\documentclass[12pt]{article}
\usepackage{amssymb}
\usepackage{epsfig}

\catcode`\@=11
\def\numberbysection{\@addtoreset{equation}{section}
        \def\theequation{\thesection.\arabic{equation}}}

\def\beq{\begin{equation}}
\def\eeq{\end{equation}}
\numberbysection
\begin{document}
\begin{titlepage}
\begin{center}
\hfill DFF  2/11/05 \\
\vskip 1.in {\Large \bf Non-abelian instantons on a fuzzy
four-sphere} \vskip 0.5in P. Valtancoli
\\[.2in]
{\em Dipartimento di Fisica, Polo Scientifico Universit\'a di Firenze \\
and INFN, Sezione di Firenze (Italy)\\
Via G. Sansone 1, 50019 Sesto Fiorentino, Italy}
\end{center}
\vskip .5in
\begin{abstract}
We study the compatibility between the $BPST \ SU(2)$ instanton and
the fuzzy four-sphere algebra. By using the projective module point
of view as an intermediate step, we are able to identify a
non-commutative solution of the matrix model equations of motion
which minimally extends the $SU(2)$ instanton solution on the
classical sphere $S^4$. We also propose to extend the non-trivial
second Chern class with the five-dimensional noncommutative
Chern-Simons term.
\end{abstract}
\medskip
\end{titlepage}
\pagenumbering{arabic}
\section{ Introduction }

Between all the known noncommutative varietes, the fuzzy four-sphere
deformation has attracted our attention because it is the only one
that incorporates the Kaluza-Klein mechanism in an elegant and
mathematically clean way \cite{1}.

As explicit application we have devoted this work to the study of
the topologically non-trivial configurations on a fuzzy four-sphere,
having as classical limit the $SU(2)$ BPST instantons on $S^4$.

The presence of the extra coordinates ( $ w_{\mu\nu} $ ) complicates
the classical limit and requires selecting between those
topologically non-trivial configurations admitting $ w_{\mu\nu} $ -
decoupling from those for which this is not possible.

The projective module point of view clarifies the whole picture
\cite{2}-\cite{3}-\cite{4}-\cite{5}-\cite{6}-\cite{7}-\cite{8}. In
fact at the $ U(1) $ level we can conceive noncommutative projectors
\cite{5} associated, through a simple link with matrix models
\cite{1}-\cite{9}-\cite{10}-\cite{11}, to reducible representations
of the fuzzy four-sphere algebra, but in this case the $ w_{\mu\nu}
$ decoupling is impossible because there is no analogous projector
in the commutative limit \cite{12}-\cite{13}.

That's why we have concentrated our attention, in this work, to the
non-abelian case, where some classical projectors exists \cite{2} (
describing the BPST instantons of the gauge group $SU(2)$ ) on which
the noncommutative deformation can be based on.

We show that it is possible to deform the classical instantons,
allowing a dependence on the extra coordinates ( $ w_{\mu\nu} $ ),
weighted by a noncommutative damping factor $ \rho \simeq
\frac{1}{N} $, which explains their decoupling in the $ N
\rightarrow \infty $.

With a careful analysis of the link between projectors and matrix
models it is possible to identify which connections do admit as a
classical limit the BPST instantons. They have a structure similar
to the 't Hooft-Polyakov monopoles connections on a fuzzy sphere,
see ref. \cite{15}. Moreover the 4d analysis is quite similar to the
2d case, apart from a fundamental passage, requiring the specific
structure of the Hopf fibration $ \pi : S^7 \rightarrow S^4 $, which
is an intrinsic 4d property.

It is also necessary extending the gauge group from $ U(2) $ to $
U(4) $ because of the dimensionality of the fuzzy four-sphere
representations. In the last part of this work we suggest how to
build a candidate for extending the second Chern class, the only
nontrivial one at a classical level.

It is convenient to take the 5d Chern-Simons term and we find
perfect agreement with the classical limit, if we compare the
noncommutative connection with the background of a $U(2)$ gauge
theory ( instead of $ U(4) $ ). These last results are still ' under
construction '  since there is a lack of a deeper mathematical
understanding of the intrinsic topological meaning of such
configurations.

\section{Review of the fuzzy four-sphere}

The fuzzy four-sphere is a non-commutative manifold, defined by the
following two general conditions

\beq \epsilon^{\mu\nu\lambda\rho\sigma} \hat{x}_\mu \hat{x}_\nu
\hat{x}_\lambda \hat{x}_\rho = C \hat{x}_\sigma \ \ \ \ \ \
\hat{x}_\mu \hat{x}_\mu = R^2 \label{21} \eeq

where $R$ is the radius of the sphere. One of these two constraints
becomes trivial in the classical limit. Both conditions can be
solved by introducing some auxiliary finite matrices $\hat{G}_\mu$
as follows

\beq \hat{x}_\mu = \rho \hat{G}_\mu \label{22} \eeq

where $\hat{G}_\mu$ is the $N$-fold symmetrized tensorial product of
the gamma matrices, see ref. \cite{1} for details.

The irreducible representations of $\hat{G}_\mu$ are labeled by this
number $N$, from which the dimension $ d_N $ of the representation $
\tilde{G}_\mu^{(N)} $ is determined to be :

\beq d_N = \frac{ ( N+1 ) ( N+2 ) ( N+3 ) }{6} \label{23} \eeq

The $C$ constant depends of course on the integer $N$. By adding the
auxiliary matrices $\hat{G}_{\mu\nu}^{(N)} = \frac{1}{2} [
\hat{G}_\mu , \hat{G}_\nu ]$ we can define a closed algebra (
$SO(5,1)$ ), with the following commutation rules

\begin{eqnarray}
& \ & [ \hat{G}^{(N)}_{\mu} , \hat{G}^{(N)}_{\nu\lambda} ] =  2 (
\delta_{\mu\nu} \hat{G}^{(N)}_{\lambda} - \delta_{\mu\lambda}
\hat{G}^{(N)}_{\nu} ) \nonumber \\
& \ & [ \hat{G}^{(N)}_{\mu\nu} , \hat{G}^{(N)}_{\lambda\rho} ] =  2
( \delta_{\nu\lambda} \hat{G}^{(N)}_{\mu\rho} + \delta_{\mu\rho}
\hat{G}^{(N)}_{\nu\lambda} - \delta_{\mu\lambda}
\hat{G}^{(N)}_{\nu\rho} - \delta_{\nu\rho}
\hat{G}^{(N)}_{\mu\lambda} ) \label{24}
\end{eqnarray}

This means in practice that we must extend the set of basic
coordinates of the four-sphere from five to fifteen:

\beq \hat{x}_\mu = \rho \hat{G}^{(N)}_\mu \ \ \ \ \ \
\hat{w}_{\mu\nu} = i \rho \hat{G}^{(N)}_{\mu\nu} = \frac{ i \rho
}{2} [ \hat{G}^{(N)}_\mu, \hat{G}^{(N)}_\nu ] \label{25} \eeq

On the fuzzy four-sphere the non-commutative coordinates are
constrained to be:

\begin{eqnarray}
& \ & [ \hat{x}_\mu, \hat{x}_\nu ] = - 2 i \rho \hat{w}_{\mu\nu}
\nonumber \\
& \ & \epsilon^{\mu\nu\lambda\rho\sigma} \hat{w}_{\mu\nu}
\hat{w}_{\lambda\rho} = - \rho ( 8 N + 16 ) \hat{x}_{\sigma}
\label{26} \end{eqnarray}

The classical sphere $S^4$ is reobtained from the large $N$ limit
keeping fixed the radius of the sphere $R$ ( $\rho \rightarrow 0$ ).
Under such limit all these coordinates become commutative, but this
doesn't mean that the extra coordinates $w_{\mu\nu}$ are decoupled
from the dynamics. Such a property must be imposed as a dynamical
request ( see later the discussion of the classical limit of the
instantons ).

The algebra (\ref{24}) satisfies several interesting relations that
can be summarized as follows:

\begin{eqnarray}
& \ & \hat{G}^{(N)}_\mu \hat{G}^{(N)}_\mu = N ( N+4 ) = c \nonumber
\\ & \ & \epsilon^{\mu\nu\lambda\rho\sigma} \hat{G}^{(N)}_\mu \hat{G}^{(N)}_\nu
\hat{G}^{(N)}_\lambda \hat{G}^{(N)}_\rho =
\epsilon^{\mu\nu\lambda\rho\sigma} \hat{G}^{(N)}_{\mu\nu}
\hat{G}^{(N)}_{\lambda\rho} = ( 8 N + 16 ) \hat{G}^{(N)}_\sigma
\label{27}
\end{eqnarray}

from which we can fix the constant $C$ and the parameter $\rho$ as

\begin{eqnarray}
& \ & C = ( 8 N + 16 ) \rho^3 \nonumber \\
& \ & R^2 = \rho^2 N ( N+4 ) \label{28}
\end{eqnarray}

We can also add the following relations

\begin{eqnarray}
& \ & \hat{G}^{(N)}_{\mu\nu} \hat{G}^{(N)}_\nu = 4 \hat{G}^{(N)}_\mu
\nonumber \\
& \ & \hat{G}^{(N)}_{\mu\nu} \hat{G}^{(N)}_{\nu\mu} = 4 N( N+4 ) =
4 c \nonumber \\
& \ & \hat{G}^{(N)}_{\mu\nu} \hat{G}^{(N)}_{\nu\lambda} = c
\delta_{\mu\lambda} +  \hat{G}^{(N)}_\mu  \hat{G}^{(N)}_\lambda - 2
\hat{G}^{(N)}_\lambda \hat{G}^{(N)}_\mu \label{29} \end{eqnarray}

The presence of the extra coordinates $\hat{w}_{\mu\nu}$ can be
understood by asserting that a fuzzy two-sphere is attached to every
point of the fuzzy four-sphere. In fact, we can take a
representation of the algebra (\ref{24}) in which the matrix $
\hat{x}_5 = \rho \hat{G}_5 $ is diagonal. Then it exists a
subalgebra $SU(2) \times SU(2)$ generated by $\hat{G}_{\mu\nu} (
\mu, \nu = 1, .., 4 )$ of the full $SO(5)$ algebra that commutes
with $\hat{x}_5$. This subalgebra can be put in a diagonal form as
follows:

\begin{eqnarray}
& \ & [ \hat{N}_i , \hat{N}_j ] = i \epsilon_{ijk} \hat{N}_k
\nonumber \\
& \ & [ \hat{M}_i , \hat{M}_j ] = i \epsilon_{ijk} \hat{M}_k
\nonumber \\
& \ & [ \hat{N}_i , \hat{M}_j ] = 0 \label{210} \end{eqnarray}

where $\hat{N}_i$ and $\hat{M}_i$ are appropriate combinations of
$\hat{G}_{\mu\nu} ( \mu, \nu = 1,...4 )$, see ref. \cite{1}. The
Casimir of each $SU(2)$ algebra can be computed in terms of the
eigenvalue $G_5$ of the matrix $\hat{x}_5$ :

\begin{eqnarray}
& \ &  \hat{N}_i \hat{N}_i = \frac{1}{16} ( N + G_5 ) ( N + 4 + G_5
) \nonumber \\
& \ &  \hat{M}_i \hat{M}_i = \frac{1}{16} ( N - G_5 ) ( N + 4 - G_5
) \label{211} \end{eqnarray}

When the eigenvalue $G_5$ takes its maximum value $N$, i.e. we stay
on top of the fuzzy four-sphere, one $SU(2)$ algebra decouples:

\beq \hat{N}_i \hat{N}_i = \frac{ N ( N+2 )}{4} \ \ \ \ \ \ \
\hat{M}_i \hat{M}_i = 0 \label{212} \eeq

and therefore we can conclude that there is only one fuzzy
two-sphere attached to the north pole of the fuzzy four-sphere. By
using the $SO(5)$ symmetry we can extend this result to every point
of the fuzzy four-sphere.

Such extra fuzzy two-sphere is a kind of internal space or spin,
which of course complicates the analysis of a field theory defined
on a fuzzy four-sphere.

Gauge theories can be defined by considering the following matrix
model

\beq S = - \frac{1}{\rho^2} Tr \left[ \frac{1}{4} [ X_\mu, X_\nu ] [
X_\mu, X_\nu ] + \frac{k}{5} \epsilon^{\mu\nu\lambda\rho\sigma}
X_\mu X_\nu X_\lambda X_\rho X_\sigma \right] \label{213} \eeq

where the indices $ \mu, \nu , .., \sigma $ take the values $ 1,..,
5$ and are contracted with the Euclidean metric, while
$\epsilon^{\mu\nu\rho\lambda\sigma}$ is the ( $SO(5)$ invariant )
totally antisymmetric tensor. $X_\mu$ are $ d_N \times d_N $
hermitian matrices ( when $d_N$ is defined in eq. (\ref{23}) ), and
$ k $ is a dimensional constant depending on $N$. The second term is
known as Myers term and it can be understood in terms of branes
\cite{11}.

The gauge symmetry is defined through the following unitary symmetry

\beq X_\mu = U^{\dagger} X_\mu U \ \ \ \ \ \ \ \ \ \ U U^{\dagger} =
U^{\dagger} U = 1 \label{214} \eeq

The $k$ constant is determined by the condition that the matrix
model admits as a classical solution the fuzzy four-sphere (
\ref{21} ):

\begin{eqnarray}
& \ & X_\mu = \hat{x}_\mu = \rho \hat{G}_\mu^{(N)} \nonumber \\
& \ & [ X_\nu, [ X_\mu, X_\nu ]] + k
\epsilon^{\mu\nu\lambda\rho\sigma} X_\nu X_\lambda X_\rho X_\sigma =
0 \label{215} \end{eqnarray}

Then $k$ can be identified with

\beq k = \frac{2}{\rho ( N+2 )} \label{216} \eeq

It is also possible to define an action of Yang-Mills with mass term
having the fuzzy four-sphere as classical solution:

\beq S = - \frac{1}{g^2} Tr \left[ \frac{1}{4} [ X_\mu, X_\nu ] [
X_\mu, X_\nu ] + 8 \rho^2 X_\mu X_\mu \right] \label{217} \eeq

The more general action can be identified as a linear combination of
two basic actions ( \ref{213} ) and ( \ref{217} )

\begin{eqnarray}
S(\lambda) & = & - \frac{1}{\rho^2} Tr \left[ \frac{1}{4} [ X_\mu,
X_\nu ] [ X_\mu, X_\nu ] + \right. \nonumber \\
& + & \left. \frac{2 \lambda}{5 ( N+2 ) \rho}
\epsilon^{\mu\nu\lambda\rho\sigma} X_\mu X_\nu X_\lambda X_\rho
X_\sigma + 8 ( 1 - \lambda ) \rho^2 X_\mu X_\mu \right] \label{218}
\end{eqnarray}

The term proportional to $\lambda$ is the analogue of the
five-dimensional Chern-Simons term which will be useful in the final
discussion as a candidate for extending the second Chern class.

The construction of a non-commutative gauge theory on a fuzzy
four-sphere is completed by developing the generic hermitian
matrices $X_\mu$ around the classical background $\hat{x}_\mu$ :

\beq X_\mu = \hat{x}_\mu + \rho R \hat{a}_\mu \label{219} \eeq

Since we are in a sort of Kaluza-Klein theory, when we develop the
$d_N \times d_N$ fluctuation matrix $\hat{a}_\mu$, we must take into
account that the field depends not only on the coordinates of the
sphere, but also on the extra coordinates $\hat{w}_{\mu\nu}$. This
requires to extend the basis of the functional space from the usual
spherical harmonics to irreducible representations of $SO(5)$, which
are labeled by two parameters $r_1, r_2$ with $ 0 \leq r_2 \leq r_1
$. The condition to be on a fuzzy four-sphere implies a truncation
of this functional space, which requires that the principal
parameter $ r_1 $ is constrained to be $ r_1 \leq N $ ( for higher
dimensional fuzzy spheres see for example \cite{14} ).

In the particular case $ r_2 = 0 $ the spherical harmonics of the
sphere are recovered. Summarizing the general development of the
fluctuation matrix $\hat{a}$ is given by

\beq \hat{a} ( \hat{x}, \hat{w} ) =  \sum_{r_1 = 0}^{n} \sum_{r_2 =
0}^{r_1} \sum_{m_i} a_{r_1 r_2 m_i} \hat{Y}_{r_1 r_2 m_i} ( \hat{x},
\hat{w} ) \label{220} \eeq

where $m_i$ are the extra quantum numbers which are necessary to
label the representation.

We need also the definition of derivative operators

\begin{eqnarray}
& \ & Ad ( \hat{G}_\mu ) \rightarrow - 2i \left( w_{\mu\nu}
\frac{\partial}{\partial x_{\nu}} - x_{\nu} \frac{\partial}{\partial
w_{\mu\nu}} \right) \nonumber \\
& \ & Ad ( \hat{G}_{\mu\nu} ) \rightarrow 2 \left( x_\mu
\frac{\partial}{\partial x_{\nu}} - x_{\nu} \frac{\partial}{\partial
x_{\mu}} - w_{\mu\lambda}\frac{\partial}{\partial w_{\lambda\nu}} +
w_{\nu\lambda}\frac{\partial}{\partial w_{\lambda\mu}} \right)
\label{221} \end{eqnarray}

The integration of the classical gauge theory is replaced by the
trace in the matrix model action. This correspondence is complicated
by the presence of the $2$-dim. internal space $N_i$.

Finally the Laplacian on the sphere has two possible extension, $ Ad
( \hat{G}_{\mu\nu} )^2 $ and $ Ad ( \hat{G}_{\mu} )^2 $. The natural
choice for a matrix model, in which we develop the $X_\mu$ matrices
around the background $ \hat{x}_\mu = \rho \hat{G}_\mu $, is given
by $ Ad ( \hat{G}_{\mu} )^2 $, that in the $ w_{\mu\nu} \rightarrow
0 $ limit reproduces the usual Laplacian on the classical
four-sphere. The action of these two Laplacians on the spherical
harmonics is as follows:

\begin{eqnarray}
& \ & \frac{1}{4} [ \hat{G}_\mu, [ \hat{G}_\mu, \hat{Y}_{r_1,r_2} ]]
= ( r_1 ( r_1 + 3 ) - r_2 ( r_2 +1 )) \hat{Y}_{r_1,r_2} \nonumber \\
& \ & - \frac{1}{8} [ \hat{G}_{\mu\nu}, [ \hat{G}_{\mu\nu},
\hat{Y}_{r_1,r_2} ]] = ( r_1 ( r_1 + 3 ) + r_2 ( r_2 +1 ))
\hat{Y}_{r_1,r_2} \label{222} \end{eqnarray}

\section{U(2) projectors and BPST instantons}

In our article \cite{5}  we postulated some $U(1)$ non-commutative
projectors, but then we realized that in this case the extra
coordinates do not decouple in the classical limit.

Hence to reach a physically meaningful result it is necessary to
study at a non-commutative level the case of $SU(2)$ instantons,
classically described in terms of projective modules in ref.
\cite{2}.

We recall that to obtain a projective module description for a
$SU(2)$ gauge theory on the $S^4$ sphere it is necessary to make use
of the Hopf projection $ \pi : S^7 \rightarrow S^4 $ and the
quaternion field $H$. In particular the function spaces we are
interested in are $ A_H = C^{\infty} ( S^4, H ) $, the algebra of
smooth functions taking values in $H$ on the basic space $S^4$, and
$ B_H = C^{\infty} ( S^7, H ) $, the algebra of smooth functions
with values in $H$ on the total space $S^7$.

The projector, whose elements belong to $A_H$, can be built, using
the Hopf fibration $ \pi : S^7 \rightarrow S^4 $, in terms of a
vector, whose elements belong to $B_H$, and this condition is
important to assure the non-triviality of the projector and the
intrinsic topological nature of the solution.

\begin{eqnarray}
p & = & | \psi >< \psi | \ \ \ \ \ \ \ \ \ < \psi | \psi > = 1 \nonumber \\
| \psi > & = & \left( \begin{array}{c} \psi_1 \\ .. \\ .. \\ \psi_N
\end{array} \right)
\label{31} \end{eqnarray}

To realize the Hopf projection $ \pi : S^7 \rightarrow S^4 $ it is
necessary to introduce a couple of quaternions ( analogously to the
case $ \pi : S^3 \rightarrow S^2 $  which is described by a couple
of complex coordinates ) subject to the constraint

\beq S^7 = \{ ( a, b ) \in H^2 , | a |^2 + | b |^2 = 1 \} \label{32}
\eeq

on which the following right action is defined

\beq S^7 \times Sp(1) \rightarrow S^7 \ \ \ \ \ \ ( a, b) w = ( aw,
bw ) \ \ \ \ \ w \in Sp(1) \ \ \ w \overline{w} = 1 \label{33} \eeq

keeping invariant the $S^7$ sphere. In terms of the quaternions $ a,
b$ the Hopf projection $ \pi : S^7 \rightarrow S^4 $ is realized as

\begin{eqnarray}
x_1 & = & a \overline{b} + b \overline{a} \ \ \ \ \xi = a
\overline{b} - b \overline{a} = - \overline{\xi} \ \ \ \ \ x_5 = | a
|^2 - | b |^2 \nonumber \\
\sum_{\mu = 1}^5 {( x_{\mu} )}^2 & = & {( | a |^2 + | b |^2 )}^2 = 1
\label{34} \end{eqnarray}

This mapping determines what are the $Sp(1)$ invariant combinations
on $S^7$, i.e. functions taking values on $S^4$:

\begin{eqnarray}
| a |^2 & = & \frac{1}{2} ( 1 + x_5 ) \nonumber \\
| b |^2 & = & \frac{1}{2} ( 1 - x_5 ) \nonumber \\
a \overline{b} & = & \frac{1}{2} ( x_1 + \xi ) \label{35}
\end{eqnarray}

Then it is not difficult to construct, at least for instanton number
$ k =1 $ a projector, whose entries belong to the $Sp(1)$ invariant
combinations, starting from the following vector $| \psi > $:

\beq | \psi > = \left( \begin{array}{c} a \\ b \end{array} \right)
\label{36} \eeq

satisfying the normalization condition $ < \psi | \psi > = | a |^2 +
| b |^2 = 1 $ on $S^7$. We can define a projector $ p \in M_2 ( A_H
) $ as

\beq p = | \psi >< \psi | = \left( \begin{array}{cc} | a |^2 & a
\overline{b} \\ b \overline{a} & | b |^2 \end{array} \right) =
\frac{1}{2} \left( \begin{array}{cc} 1 + x_5 & x_1 + \xi \\
x_1 - \xi & 1 - x_5 \end{array} \right) \label{37} \eeq

Clearly if we transform $ | \psi > $ with right action $ Sp(1) : S^7
\times Sp(1) \rightarrow S^7 $

\beq | \psi > \rightarrow | \psi^w > = \left( \begin{array}{c} a w  \\
b w \end{array} \right) = | \psi > w \ \ \  \forall w \in Sp(1)
\label{38} \eeq

the projector $p$ remains invariant ( another way to say that its
elements belong to the $A_H$ algebra instead of $B_H$ ).

The associated Chern classes are

\begin{eqnarray}
& \ & C_1 ( p ) = - \frac{1}{ 2 \pi i} Tr ( p ( dp )^2 ) \nonumber
\\ & \ & C_2 (p) = - \frac{1}{ 8 \pi^2} [ Tr ( p (dp)^4 ) - (C_1
(p))^2 ] \label{39} \end{eqnarray}

Since the $2$-form $ p (dp)^2 $ has values in the pure imaginary
quaternions, its trace is null

\beq C_1 ( p) = 0 \label{310} \eeq

Its second Chern class is instead non-trivial

 \beq C_2 (p) = - \frac{3}{8 \pi^2} d ( vol ( S^4 ) ) \label{311}
\eeq

and it gives rise to a non-trivial topological number ( Chern number
)

\beq c_2 (p) = \int_{S^4} C_2 (p) = - \frac{3}{8\pi^2} \int_{S^4} d
( vol ( S^4 )) = - \frac{3}{8\pi^2} \frac{8\pi^2}{3} = - 1
\label{312} \eeq

The $1$-form connection, associated with the projector $p$

\beq A^{\nabla} = < \psi | d | \psi > = \overline{a} d a +
\overline{b} d b \label{313} \eeq

is anti-hermitian taking values in the purely imaginary quaternions,
that can be identified with the Lie algebra $Sp(1) \sim SU(2)$.

The non-trivial moduli space of the $k=1$ instanton is generated by
the elements $ g \in SL(2,H) $ ( belonging to the conformal group ),
acting on the left:

\beq | \psi > \rightarrow | \psi^g > = \frac{1}{ [ < \psi
|g^{\dagger} g | \psi > ]^{\frac{1}{2}}  } g | \psi > \label{314}
\eeq

The quotient of $SL(2,H)$ by the trivial subgroup $Sp(2)$ generates
a five parameter family of the $ k = 1 $ $SU(2)$ instantons.

At a noncommutative level, the existence of a classical projector
helps in defining a corresponding non-commutative projector, whose
elements necessary belong to the whole fuzzy four-sphere algebra,
but where the dependence on the extra coordinates $
\hat{w}_{\mu\nu}$ is controlled by a factor $\rho$ ( $ \rho \simeq
\frac{1}{N} $ ). This assures the necessary decoupling of the extra
coordinates in the $\rho \rightarrow 0$ limit, as physically
required \cite{13}.

To reach such property we must rewrite the Hopf fibration $ \pi :
S^7 \rightarrow S^4 $ in terms of four complex coordinates:

\begin{eqnarray}
& \ & x_1 = \rho ( \alpha_1 + \overline{\alpha}_1 ) \ \ \ \ \ \ x_2
= i \rho ( \alpha_1 - \overline{\alpha}_1 ) \nonumber \\
& \ & x_3 = \rho ( \alpha_2 + \overline{\alpha}_2 ) \ \ \ \ \ \ x_4
= i \rho ( \alpha_2 - \overline{\alpha}_2 ) \nonumber \\
& \ & x_5 = \rho ( a_0 \overline{a}_0 + a_1 \overline{a}_1 - a_2
\overline{a}_2 - a_3 \overline{a}_3 ) \nonumber \\
& \ & \alpha_1 = a_0 \overline{a}_2 + a_3 \overline{a}_1 \ \ \ \ \ \
\alpha_2 = a_0 \overline{a}_3 - a_2 \overline{a}_1 \nonumber \\
& \ & \sum_i x^2_i = \rho^2 \ \ \ \ \ \ \ \sum_i | a_i |^2 = 1
\label{315} \end{eqnarray}

In ref . \cite{5} we noticed that quantizing such complex
coordinates and keeping the same mapping (\ref{315})

\beq [ a_i, a^{\dagger}_j ] = \delta_{ij} \ \ \ \ \  [ a_i, a_j ] =
0 \label{316} \eeq

the resulting $\hat{x}_i$ coordinates are part of an algebra,
coinciding with the fuzzy four-sphere algebra. In fact we can show
that the total number operator

\beq \hat{N} = a^{\dagger}_0 a_0 + a^{\dagger}_1 a_1 + a^{\dagger}_2
a_2 + a^{\dagger}_3 a_3 \label{317} \eeq

has as eigenvalue $N$ on a irreducible representation of the
algebra, and the Casimir $\hat{x}_i^2$ can be re-expressed in terms
of the total number operator $\hat{N}$:

\beq \sum_i \hat{x}^2_i = \rho^2 \hat{N} ( \hat{N} + 4 ) = R^2
\label{318} \eeq

In terms of these oscillators, the non-commutative analogues of the
quaternions $ ( a, b ) $ can be expressed as

\beq a = \left( \begin{array}{cc} a_0 & - a^{\dagger}_1 \\
a_1 & a^{\dagger}_0 \end{array} \right) \ \ \ \ \ \
b = \left( \begin{array}{cc} a_2 & - a^{\dagger}_3 \\
a_3 & a^{\dagger}_2 \end{array} \right) \label{319} \eeq

from which it is clear that the combination $ a \overline{b} $ is
inside the fuzzy four-sphere algebra, while $ \overline{b} a $ is
outside. The only combinations belonging to the algebra are of the
type ( $ a \overline{a}, a \overline{b}, b \overline{a}, b
\overline{b} $ ), which impose the following guess for the vector $
| \psi_0 > $

\beq | \psi_0 > = \left( \begin{array}{c} a \\ b \end{array} \right)
\label{320} \eeq

By imposing the normalization condition

\beq < \psi_0 | \psi_0 > = ( \overline{a}  \ \overline{b} ) \left(
\begin{array}{c} a \\ b \end{array} \right) = \overline{a} a +
\overline{b} b = \left( \begin{array}{cc} \hat{N} & 0 \\ 0 &
\hat{N+4} \end{array} \right) \label{321} \eeq

we obtain a diagonal matrix whose elements depend only on the total
number operator $\hat{N}$. Redefining $ | \psi_0 > $ as

\beq | \psi_0 > \rightarrow | \psi > = \left( \begin{array}{c} a' \\
b' \end{array} \right) \ \ \ \ \ \ a' = a \sqrt{ h( \hat{N} ) } \ \
\ b' = b \sqrt{ h( \hat{N} ) } \label{322} \eeq

with the position

\beq h( \hat{N} ) = \left( \begin{array}{cc} \frac{1}{\hat{N}} & 0 \\
0 & \frac{1}{\hat{N+4}} \end{array} \right) \label{323} \eeq

we can develop the resulting projector as

 \beq p_n = | \psi >< \psi | = \left(
\begin{array}{c} a \\ b \end{array} \right) \left( \begin{array}{cc}
h(\hat{N}) & 0 \\ 0 & h(\hat{N})
\end{array} \right) ( \overline{a}  \overline{b} ) =
\left( \begin{array}{cc} a h(\hat{N}) \overline{a} & a h(\hat{N})
\overline{b} \\ b h(\hat{N}) \overline{a} & b h(\hat{N})
\overline{b}
\end{array} \right) \label{324} \eeq

We notice that its elements are functions also of the extra
coordinates $\hat{w}_{\mu\nu}$. However, differently from the $U(1)$
case, the dependence on the extra coordinates has a factor $ \rho
\sim \frac{1}{N} $, that assures their decoupling in the classical
limit. We can write explicitly its first entry

\begin{eqnarray}
& \ & a h(\hat{N}) \overline{a} =  \left( \begin{array}{cc} a_0 & - a^{\dagger}_1 \\
a_1 & a^{\dagger}_0 \end{array} \right) \left( \begin{array}{cc}
\frac{1}{\hat{N}} & 0 \\ 0 & \frac{1}{\hat{N}+4} \end{array} \right)
\left( \begin{array}{cc} a^{\dagger}_0 &
 a^{\dagger}_1 \\
- a_1 & a_0 \end{array} \right) = \nonumber \\
& \ & = \left( \begin{array}{cc} \frac{1}{\hat{N}+1} a_0
a_0^{\dagger} + \frac{1}{\hat{N}+3} a_1^{\dagger} a_1 & \left(
\frac{1}{\hat{N}+1} - \frac{1}{\hat{N}+3} \right) a_0 a_1^{\dagger}
\\ \left( \frac{1}{\hat{N}+1} - \frac{1}{\hat{N}+3} \right) a_1
a_0^{\dagger} & \frac{1}{\hat{N}+1} a_1 a_1^{\dagger} +
\frac{1}{\hat{N}+3} a_0^{\dagger} a_0
\end{array}\right) \label{325} \end{eqnarray}

At this level we can substitute $\hat{N}$ with its eigenvalue $N$,
making $p_N$ a projector of the fuzzy four-sphere. The dependence on
the extra coordinates is proportional to $ \left( \frac{1}{N+1} -
\frac{1}{N+3} \right) $, and therefore is vanishing for $ N
\rightarrow \infty $. The trace of the projector can be computed as

\begin{eqnarray} Tr \ p_N & = & \left( \frac{N+4}{N+1} + \frac{N}{N+3}\right) Tr \ I =
\nonumber \\ &  = & \left( \frac{(N+2)(N+3)(N+4)}{6} + \frac{N
(N+1)(N+2)}{6}\right) = \nonumber \\
& = &  2 Tr \ I + ( N+2 ) < Tr \ 1_P = 4 Tr \ I \label{326}
\end{eqnarray}

and it is always an integer, for every value of $N$. There are no
constraints on the possible values of $N$.

\section{Connection with matrix models}

As discussed in \cite{15}, the connection between projectors and
matrix models in the non-abelian case is not as simple as in the
$U(1)$ case, however it is still convenient starting from the
following guess

\beq X_i = < \psi | G_i | \psi > \label{41} \eeq

and correct this formula later on to reach a solution of the
non-commutative equations of motion.

By looking at the form of $ | \psi > $

\beq | \psi > = \left( \begin{array}{c} \left(\begin{array}{cc} a_0
& - a^{\dagger}_1 \\ a_1 & a^{\dagger}_0 \end{array} \right)
\\ \left( \begin{array}{cc} a_2 & - a^{\dagger}_3 \\ a_3 & a^{\dagger}_2 \end{array}
\right) \end{array} \right) f( \hat{N} ) \ \ \ \ \ f(\hat{N}) =
\left( \begin{array}{cc} \frac{1}{\sqrt{\hat{N}}} & 0 \\ 0 &
\frac{1}{\sqrt{\hat{N}+4}} \end{array} \right) \label{42} \eeq

it is not a priori obvious that the expectation value $ < \psi | G_i
| \psi > $ reduces to a diagonal form. Since this really happens, we
believe that such property is another consequence of the underlying
Hopf projection $ \pi : S^7 \rightarrow S^4 $.

Let us check how it happens in details. The expectation value

\begin{eqnarray} < \psi | G_i | \psi > &  = & \left( \begin{array}{cc} \sum_k
a^{\dagger}_k G_i a_k & - a^{\dagger}_0 G_i a^{\dagger}_1 +
a^{\dagger}_1 G_i a^{\dagger}_0 - a^{\dagger}_2 G_i a^{\dagger}_3 +
a^{\dagger}_3 G_i a^{\dagger}_2
\\ a_0 G_i a_1 - a_1 G_i a_0 + a_2 G_i a_3 - a_3 G_i a_2 &
\sum_k a_k G_i a^{\dagger}_k \end{array} \right) \nonumber \\
& \ & \cdot \left(
\begin{array}{cc} \frac{1}{\hat{N}} & 0 \\ 0 & \frac{1}{\hat{N}+4}
\end{array} \right) \label{43} \end{eqnarray}

contains in principle off-diagonal terms. For example let us check
the term $ (X_i)_{12} $

\begin{eqnarray}
( X_i )_{12} & = & - a^{\dagger}_0 G_i a^{\dagger}_1 + a^{\dagger}_1
G_i a^{\dagger}_0 - a^{\dagger}_2 G_i a^{\dagger}_3 + a^{\dagger}_3
G_i a^{\dagger}_2 = \nonumber \\
& = & - a^{\dagger}_0 [ G_i ,  a^{\dagger}_1 ] + a^{\dagger}_1 [ G_i
, a^{\dagger}_0 ] - a^{\dagger}_2 [ G_i , a^{\dagger}_3 ] +
a^{\dagger}_3 [ G_i ,  a^{\dagger}_2 ] \label{44} \end{eqnarray}

In the case $ i = 5 $ we obtain

\begin{eqnarray}
& \ & G_5 = a^{\dagger}_0 a_0 +  a^{\dagger}_1 a_1 - a^{\dagger}_2
a_2
- a^{\dagger}_3 a_3 \nonumber \\
& \Rightarrow & (X_5)_{12} = - a^{\dagger}_0 a^{\dagger}_1 +
a^{\dagger}_1 a^{\dagger}_0 - a^{\dagger}_2 a^{\dagger}_3 +
a^{\dagger}_3 a^{\dagger}_2 = 0 \label{45}
\end{eqnarray}

In the other cases we can simplify the discussion by taking the
complex coordinates $ G_{1\pm} $

\begin{eqnarray}
& \ & G_{1-}  =  a_0 a^{\dagger}_2 + a_3 a^{\dagger}_1 \ \ \ \ \ \ \
\ \ \ \ \ \ \ \ \ \ \ \ \ \ G_{1+}  =  a^{\dagger}_0 a_2 +
a^{\dagger}_3 a_1
\nonumber \\
& \Rightarrow & (X_{1-})_{12} = a^{\dagger}_1 a^{\dagger}_2 -
a^{\dagger}_2 a^{\dagger}_1 = 0 \ \ \ \ \ \ \ \ (X_{1+})_{12} = -
a^{\dagger}_0 a^{\dagger}_3 + a^{\dagger}_3 a^{\dagger}_0 = 0
\label{46}
\end{eqnarray}

and $ G_{2\pm} $

\begin{eqnarray}
& \ & G_{2-}  =  a_0 a^{\dagger}_3 - a_2 a^{\dagger}_1 \ \ \ \ \ \ \
\ \ \ \ \ \ \ \ \ \ \ \ \ G_{2+}  =  a^{\dagger}_0 a_3 - a^{\dagger}_2 a_1 \nonumber \\
& \Rightarrow & (X_{2-})_{12} = a^{\dagger}_1 a^{\dagger}_3 -
a^{\dagger}_3 a^{\dagger}_1 = 0 \ \ \ \ \ \ \ \ \ (X_{2+})_{12} =
a^{\dagger}_0 a^{\dagger}_2 - a^{\dagger}_2 a^{\dagger}_0 = 0
\label{47}
\end{eqnarray}

In all cases a welcome cancellation appears, the reader can verify
it also for the case $ (X_i)_{21} $. The resulting diagonal terms
can be computed by using the commutation relations of the
oscillators

\beq < \psi | G_i | \psi > = \left( \begin{array}{cc}
\frac{\hat{N}-1}{\hat{N}} G_i & 0 \\ 0 & \frac{\hat{N}+5}{\hat{N}+4}
G_i
\end{array} \right) \label{48} \eeq

As we already discussed in our previous articles
\cite{12}-\cite{13}, the $G_i$ action on the vector $ | \psi > $
cannot be smoothly connected to ordinary derivative operators on the
sphere in the classical limit unless we project the vector $ | \psi
> $ on the fuzzy four-sphere algebra.

Since we must maintain invariant the form $U(2)$ of the projector (
which is smoothly connected to the BPST $ SU(2)$ instantons on the
sphere $S^4$ ), we correct the vector $ | \psi > $ with a
quasi-unitary operator such that $ | \psi' > =  | \psi >  U $ has
all elements belonging to the fuzzy four-sphere algebra:

\begin{eqnarray}
& \ & | \psi' > = | \psi> U \ \ \ \ \ U U^{\dagger} = 1 \nonumber \\
& \ & P = | \psi >< \psi | = | \psi' >< \psi' | \label{49}
\end{eqnarray}

This operator $U$, as in the case of non-commutative monopoles,
plays an essential role to define the non-abelian topology. However,
since the Hilbert space of the fuzzy four-sphere is generated by
four oscillators we find an unexpected difficulty in solving the
constraint $ U U ^{\dagger} = 1 $ with a $ 2 \times 2 $ matrix.

This difficulty can be overcome only by embedding the $U(2)$
projector in a larger gauge group. In fact we can solve the
constraint $ U U ^{\dagger} = 1 $ with a $ 4 \times 4 $ matrix,
which requires embedding the projector in a $ U(4) $ gauge theory.

During the discussion of the final solution, we will notice that
such difficulty is related to the dimensionality of the fuzzy
four-sphere irreducible representations, defined in eq. ( \ref{23}
).

\section{Embedding $U(2)$ in $U(4)$}

In the case of $U(1)$ projectors ( see ref. \cite{12}-\cite{13} ) we
were stuck into the problem of identifying the associated
connections for a certain class of projectors ( Serre-Swan theorem
for the noncommutative case ). In this section the answer to this
question comes out. The only obstacle, i.e. the construction of the
quasi-unitary operator $U$, can always be overcome by embedding the
projector into a larger gauge group.

In the present case solving the constraint $ U U^{\dagger} = 1 $,
with $ U $ represented by $ 4 \times 4 $ matrix is an easy task

\beq U = \left( \begin{array}{cccc} U_{11}^{\dagger} &
U_{12}^{\dagger} & U_{13}^{\dagger} & U_{14}^{\dagger}
\\ 0 & U_4 & 0 & 0 \\ 0 & 0 & 0 & 0 \\ 0 & 0 & 0 & 0 \end{array}
\right) \ \ \ \ \ \ \ U^{\dagger} = \left( \begin{array}{cccc}
U_{11} & 0 & 0 & 0
\\ U_{12} & U^{\dagger}_4 & 0 & 0 \\ U_{13} & 0 & 0 & 0 \\ U_{14} & 0 & 0 & 0 \end{array}
\right) \label{51} \eeq

and the obvious solution can be represented as:

\begin{eqnarray}
U_{11} & = & \sum_{n_1,n_2,n_3,n_4=0}^{\infty} | n_1, n_2, n_3, n_4
><
n_1 + 1, n_2, n_3, n_4 | \nonumber \\
U_{12} & = & \sum_{n_2,n_3,n_4=0}^{\infty} | n_2, n_3, n_4, 0
>< 0, n_2 + 1, n_3, n_4 | \nonumber \\
U_{13} & = & \sum_{n_3,n_4=0}^{\infty} | n_3, n_4, 0 , 0
>< 0, 0, n_3 + 1, n_4 | \nonumber \\
U_{14} & = & \sum_{n_4=0}^{\infty} | n_4, 0 , 0 , 0
>< 0, 0, 0, n_4 + 1| \nonumber \\
U_4 & = & \sum_{n_1,n_2,n_3,n_4=0}^{\infty} | n_1, n_2, n_3, n_4
>< n_1 , n_2, n_3, n_4 + 1| \label{52} \end{eqnarray}

The structure of this solution is so unique that it is practically
impossible to reduce it to a subspace of $ 2 \times 2 $ matrices.

This quasi-unitary operator satisfies the following property:

\beq U U^{\dagger} = \left(
\begin{array}{cccc}  \sum_{k=1}^4 U^{\dagger}_{1k} U_{1k} = 1 - | 0, 0, 0, 0
>< 0, 0, 0 , 0 | & 0 &
0 & 0 \\ 0 & U_4 U^{\dagger}_4 = 1 & 0 & 0 \\
0 & 0 & 0 & 0 \\
0 & 0 & 0 & 0
\end{array} \right) \label{53} \eeq

$ U U^{\dagger} $ is practically equivalent to the identity operator
$ 1_{U(2)} $, apart from an operator $ | 0, 0, 0, 0 >< 0, 0, 0, 0 |
$, whose action is annihilated by the vectors $ | \psi > $.

For the combination $ U^{\dagger} U $ we obtain, differently from
the $ 2d $ case ( see ref. \cite{15} ):

\beq U^{\dagger} U = \left(
\begin{array}{cccc}  1 & 0 & 0 & 0 \\
0 & 1 & 0 & 0 \\
0 & 0 &  \sum_{n_3, n_4} |
n_3, n_4 , 0 , 0 > <  n_3, n_4 , 0 , 0 | & 0 \\
0 & 0 & 0 & \sum_{n_4} | n_4 , 0 , 0 , 0
> < n_4 , 0 , 0 , 0 |
\end{array} \right) \label{54} \eeq

This property is strictly related to the dimensionality of the fuzzy
four-sphere irreducible representations ( see eq. ( \ref{23} )).

The explicit construction of the quasi-unitary operator requires to
extend minimally the vector $ | \psi > $, with entries belonging to
$ U(2) $ gauge group, to vectors with values in $ U(4) $. Since the
extension must be minimal, we restrict its non-trivial contribution
to a $ U(2) $ subgroup

\beq | \psi > = \left( \begin{array}{c} \left(\begin{array}{cccc}
a_0 & - a^{\dagger}_1 & 0 & 0 \\ a_1 & a^{\dagger}_0 & 0 & 0 \\ 0 &
0 & 0 & 0 \\ 0 & 0 & 0 & 0 \end{array} \right)
\\ \left( \begin{array}{cccc} a_2 & - a^{\dagger}_3 & 0 & 0
\\ a_3 & a^{\dagger}_2 & 0 & 0 \\ 0 & 0 & 0 & 0 \\ 0 & 0 & 0 & 0 \end{array}
\right) \end{array} \right) f( \hat{N} ) \label{55} \eeq

The expectation value (\ref{48}) with this new vector $ | \psi > $
is similar to the $ U(2) $ case:

\beq X_i = < \psi | G_i | \psi > = \left( \begin{array}{cccc}
\frac{\hat{N}-1}{\hat{N}} G_i & 0 & 0 & 0 \\ 0 &
\frac{\hat{N}+5}{\hat{N}+4} G_i & 0 & 0 \\ 0 & 0 & 0 & 0 \\ 0 & 0 &
0 & 0 \end{array} \right) \label{56} \eeq

Correcting the vector $ | \psi > $ with $ 4 \times 4 $ quasi-unitary
operators changes the dimension of the background representation
with a non-trivial mixing similar to the $2d$ case. Therefore we
must expect that

\beq ( U^{\dagger} X_i^{bg} U )_{N+1} = \left( \begin{array}{c|c|c}
( G_i )_{N+2} & 0 & 0 \\ \hline 0 & ( G_i )_N & 0 \\ \hline 0 & 0 &
0
\end{array} \right) \label{57} \eeq

where the division in blocks is different in the last matrix. Such
property is discussed in details in the appendix, but we give now an
indirect argument. As a consistency check of eq. (\ref{57}) we
notice that by taking the square of the background an interesting
simplification appears. In fact since it is proportional to a
Casimir of the fuzzy four-sphere algebra its trace is a calculable
number:

\begin{eqnarray}
& \ & Tr ( X_i^{bg} )_{N+1}^2 = 4 Tr (G_i)_{N+1} (G_i)_{N+1} =
\nonumber \\
& \ & = 4 ( N+1 ) ( N+5 ) Tr 1_{N+1} = \frac{2}{3} ( N+1 ) ( N+2 ) (
N+3 ) ( N+4 ) ( N+5 ) \label{58} \end{eqnarray}

Now let us take the square of equations (\ref{57}) and compare the
trace of both members. If we find agreement between these two
numbers, we have an indirect proof that eq. (\ref{57}) is correct.
Firstly we compute

\beq Tr ( U^{\dagger} ( X_i^{bg} X_i^{bg} ) U )_{N+1} = Tr (
U_1^{\dagger} ( X_i^{bg} X_i^{bg} ) U_1 )_{N+1} + Tr ( U_2^{\dagger}
( X_i^{bg} X_i^{bg} ) U_2 )_{N+1} \label{59} \eeq

where we have splitted  the quasi-unitary operator $ U $ in the
following two parts ( $ U = U_1 + U_2 $ ):

\beq  U_1 = \left( \begin{array}{cccc} U_{11}^{\dagger} &
U_{12}^{\dagger} & U_{13}^{\dagger} & U_{14}^{\dagger}
\\ 0 & 0  & 0 & 0 \\ 0 & 0 & 0 & 0 \\ 0 & 0 & 0 & 0 \end{array}
\right) \ \ \ \ \ \ U_2 = \left( \begin{array}{cccc} 0 & 0 & 0 & 0
\\ 0 & U_4 & 0 & 0 \\ 0 & 0 & 0 & 0 \\ 0 & 0 & 0 & 0 \end{array}
\right) \label{510} \eeq

This decomposition permits the following manipulations

\begin{eqnarray}
& \ & Tr ( U_1^{\dagger} ( X_i^{bg} X_i^{bg} ) U_1 )_{N+1} = ( N+2 )
( N+6 ) Tr ( U^{\dagger}_1 U_1 ) \nonumber \\
& \ & Tr ( U_2^{\dagger} ( X_i^{bg} X_i^{bg} ) U_2 )_{N+1} = N
( N+4 ) Tr ( U^{\dagger}_2 U_2 ) \nonumber \\
& \ & Tr ( U^{\dagger}_1 U_1 ) = \sum_{k=1}^4 Tr ( U_{1k}
U_{1k}^{\dagger} ) = Tr (1)_{N+1} + ( Tr (1)_{N+1} - Tr (1)_N ) + (
N+2 )
+ 1 = Tr (1)_{N+2} \nonumber \\
& \ & Tr ( U^{\dagger}_2 U_2 )  = Tr ( U^{\dagger}_4 U_4 ) = Tr
(1)_N \label{511} \end{eqnarray}

In summary we obtain

\beq Tr ( U^{\dagger} ( X_i^{bg} X_i^{bg} ) U )_{N+1}  = N(N+4) Tr
(1)_N + ( N+2 )( N+6 ) Tr (1)_{N+2} \label{512} \eeq

On the other hand, the second member of eq. (\ref{57}) gives rise to

\beq Tr \left( \begin{array}{c|c|c} (G_i)_{N+2} & 0 & 0 \\ \hline 0
& (G_i)_{N} & 0 \\ \hline 0 & 0 & 0 \end{array} \right)^2 = (G_i)_N
(G_i)_N Tr (1)_N + (G_i)_{N+2} (G_i)_{N+2} Tr (1)_{N+2} \label{513}
\eeq

which is exactly the same number.

Proceeding this way, the expectation value for $ | \psi' > = | \psi
> U $ can be explicitly computed as

\begin{eqnarray} & \ & | \psi' > = | \psi > U  \ \ \ \ \ \  P = | \psi' > < \psi' | =
| \psi > < \psi | \nonumber \\ & \ & X_i^{(0)} = < \psi' | X_i^{bg}
| \psi'
> = \left( \begin{array}{c|c|c} \frac{\hat{N}}{\hat{N}+1}
(G_i)_{N+2} & 0 & 0 \\ \hline 0 &
\frac{\hat{N}+4}{\hat{N}+3}(G_i)_{N} & 0 \\
\hline 0 & 0 & 0
\end{array} \right) \label{514} \end{eqnarray}

This block diagonal matrix is still not an explicit solution of the
non-commutative equations of motion. However as in the $2d$ case,
the nearest solution is easy to reach, by redefining the relation
between $ X_i $ and the vector $ | \psi' > $ in the following way:

\begin{eqnarray}
& \ & X_i^{tot} = \left( \begin{array}{cccc} f_{+} &  &  &  \\  & f_{-} & & \\
& & 0 & \\ & & & 0 \end{array} \right) X_i^{(0)} \left(
\begin{array}{cccc} f_{+} &  &  &  \\  & f_{-} & & \\ & & 0 & \\ & & & 0
\end{array}
\right) = \nonumber \\
& \ & = \left( \begin{array}{c|c|c}  (G_i)_{N+2} & 0 & 0
\\ \hline 0 & (G_i)_{N} & 0 \\ \hline 0 & 0 & 0
\end{array} \right) \label{515} \end{eqnarray}

The unknown constants $ f_{\pm} $ are constrained to be

\begin{eqnarray}
\left\{
\begin{array}{c} ( f_{+} + f_{-} )^2 = \frac{N+1}{N}
\\ ( f_{+} - f_{-} )^2 = \frac{N+1}{N+4} \end{array}
\right. \ \ \ \ \ \ \ \left\{
\begin{array}{c} f_{+} = \frac{1}{2}
\left( \sqrt{\frac{N+1}{N}} + \sqrt{\frac{N+1}{N+4}} \right)
\\  f_{-} = \frac{1}{2} \left( \sqrt{\frac{N+1}{N}} -
\sqrt{\frac{N+1}{N+4}}  \right) \end{array} \right. \label{516}
\end{eqnarray}

In conclusion, the compatibility of the $ SU(2) $ BPST instanton
with the non-commutative structure of the fuzzy four-sphere requires
two non-trivial steps:

- a sort of Kaluza-Klein extension of the coordinates of the
classical sphere from five to fifteen coordinates;

- the extension of the gauge group from $ SU(2) $ to $ U(4) $, not
expected from the projector point of view;

We have only an indirect check that this non-commutative solution is
a smooth deformation of the classical $ SU(2) $ BPST instanton,
however we believe that this is the minimal way to realize such an
extension.

As a last remark, we want to comment on the possibility to define a
topological number, which extends the non-trivial second Chern class
of BPST instantons. As a non-commutative deformation we suggest to
choose the 5-dimensional Chern-Simons term:

\beq S_{ch} = - \frac{5 \beta}{12} \ Tr \left[ \frac{1}{5 ( N+2 )
\rho } \epsilon^{\mu \nu \lambda \rho \sigma} X_\mu X_\nu X_\lambda
X_\rho X_\sigma - 4 \rho^2 X_\mu X_\mu \right] \label{517} \eeq

Since the solution, defined in formula (\ref{515}), is a ( reducible
) representation of the fuzzy four-sphere algebra, the evaluation of
$S_{CS}$ on it reduces to the evaluation of the following simplified
action

\beq S_m = \beta Tr ( X_\mu X_\mu ) \label{518} \eeq

To obtain a well-defined result, we must compare the non-commutative
solution

\begin{eqnarray} S_m ( X_i^{tot} ) & = & \beta \rho^2 \ [ Tr ( G_i G_i )_{N+2} +
Tr ( G_i G_i )_{N} ] = \nonumber \\
& = & \beta \rho^2 \ Tr (1)_{N+1} [ N ( N+1 ) + ( N+5 ) ( N+6 ) ] = \nonumber \\
& = & 2 \beta \rho^2 \ Tr (1)_{N+1} [ N^2 + 6 N + 15 ] \label{519}
\end{eqnarray}

with the background of a $U(2)$ gauge theory  ( and not $U(4)$ )

\begin{eqnarray}
S_m ( X_i^{bg}(U(2)) ) & = & 2 \beta \rho^2 \ Tr ( G_i G_i )_{N+1} =
\nonumber \\
& = & 2 \beta \rho^2 Tr (1)_{N+1} ( N+1 )( N+5 ) = 2 \beta \rho^2 Tr
(1)_{N+1} ( N^2 + 6N + 5 ) \nonumber \\
& \ & \label{520} \end{eqnarray}

as it is clear from the following calculation:

\beq S_m ( X_i^{tot} ) - S_m ( X_i^{bg}(U(2)) ) = 20 \beta \rho^2 \
Tr ( 1 )_{N+1} \label{521} \eeq

Since all the dependence on $N$ is contained in the term $ Tr( 1
)_{N+1} $, it is enough to renormalize the trace of the Chern-Simons
term with $ Tr( 1 )_{N+1} $, to obtain an integer number, which
probably coincides with the classical topological number, under the
hypothesis that such limit is smooth. This analysis is however
probably incomplete since there is an ambiguity of adding constant
terms to eq. (\ref{515}) or to the background to reach the classical
limit ( see the final discussion of non-commutative monopoles in
ref. \cite{15} ). Solving such ambiguity requires a deeper
mathematical comprehension of the results of this paper.

\section{Conclusions}

In this work we have succeeded in showing that our method based on
projective modules and matrix models can be extended to the
non-abelian case in 4d. In particular the extension of the
instantons to the fuzzy four-sphere requires two important steps:

i) the Kaluza-Klein mechanism, which extends the five coordinates of
the $S^4$ sphere to the fifteen ones of the noncommutative case. It
should be noted that the necessary decoupling of the extra
coordinates $ w_{\mu\nu} $ is assured directly at the projector
level;

ii) the extension of the gauge group from $SU(2)$ to $U(2)$ for what
concerns the projector and to $U(4)$ for the connection. Such
distinction is a consequence of the dimensionality of the fuzzy-four
sphere representations.

In the last part of the work, we suggest how to extend the second
nontrivial Chen class of the instanton to the $5$-dim Chern-Simons
action. We find agreement with the existence of a topological number
also at a noncommutative level, although our proposal must be
supported by more rigorous mathematical arguments.

\appendix
\section{Appendix}

The quasi-unitary operator $ U $ defined in eqs. (\ref{51}) and (
\ref{52}) , when acting on the background, induces a non-trivial
non-commutative topology, extension of the classical $ SU(2) $ BPST
instanton. In this appendix we show how to check the main property,
which is used in eq. ( \ref{57} ) and ( \ref{514} )

\beq ( U^{\dagger} X_i^{bg} U )_{N+1} = \left( \begin{array}{c|c|c}
( G_i )_{N+2} & 0 & 0 \\ \hline 0 & ( G_i )_N & 0 \\ \hline 0 & 0 &
0
\end{array} \right) \label{a1} \eeq

First we notice that it is worth separating $ U $ in two pieces, $ U
= U_1 + U_2 $, as in eq. ( \ref{510} ), and studying the two cases
separately:

\begin{eqnarray}
( U^{\dagger}_1 X_i^{bg} U_1 )_{N+1} = \left( \begin{array}{c|c|c}
(G_i)_{N+2} & 0 & 0 \\ \hline 0 & 0 & 0 \\ \hline 0 & 0 & 0
\end{array} \right) \nonumber \\
( U^{\dagger}_2 X_i^{bg} U_2 )_{N+1} = \left( \begin{array}{c|c|c} 0
& 0 & 0 \\ \hline 0 & (G_i)_N & 0 \\ \hline 0 & 0 & 0
\end{array} \right)
\label{a2} \end{eqnarray}

The second case is rather evident and we concentrate ourself only on
the $U_1$ case

\beq ( U^{\dagger} X_i^{bg} U )_{N+1} = \left( \begin{array}{cccc}
U_{11} G_i U^{\dagger}_{11} & U_{11} G_i U^{\dagger}_{12} & U_{11}
G_i U^{\dagger}_{13} & U_{11} G_i
U^{\dagger}_{14} \\
U_{12} G_i U^{\dagger}_{11} & U_{12} G_i U^{\dagger}_{12} & U_{12}
G_i U^{\dagger}_{13} & U_{12} G_i
U^{\dagger}_{14} \\
U_{13} G_i U^{\dagger}_{11} & U_{13} G_i U^{\dagger}_{12} & U_{13}
G_i
U^{\dagger}_{13} & U_{13} G_i U^{\dagger}_{14} \\
U_{14} G_i U^{\dagger}_{11} & U_{14} G_i U^{\dagger}_{12} & U_{14}
G_i U^{\dagger}_{13} & U_{14} G_i U^{\dagger}_{14} \end{array}
\right) \label{a3} \eeq

Since this case is rather intricate, we firstly check the dimension
of this representation. When restricting $ U^{\dagger} X_i^{bg} U $
to a total fixed number $ \sum  n_i  = N+1 $, the following
properties are clear:

1) $ (U_{11} G_i U^{\dagger}_{11})_{N+1} $ is a square matrix,
having the same dimension of the $ (G_i)_{N+1} $ representation,
i.e. $ d_{N+1} $;

2) in the case $ (U_{11} G_i U^{\dagger}_{12})_{N+1} $, its elements
$ \neq 0 $ are concentrated into a rettangular matrix with short
side $ \frac{ ( N+2 ) ( N+3 ) }{2} $ ;

3) in the case $ (U_{11} G_i U^{\dagger}_{13})_{N+1} $ the short
side is $ ( N+2 ) $;

4) finally $ (U_{11} G_i U^{\dagger}_{14})_{N+1} $ can be reduced to
a single column.

In summary we have the following structure of the representation
(\ref{a3}) , ( adding for completeness the term due to $U_2$ ):

\beq \left(
\begin{array}{c||c|c||c|c||c|c} U_{11} G_i U_{11}^{\dagger} & U_{11} G_i
U_{12}^{\dagger} & 0 & U_{11} G_i U_{13}^{\dagger} & 0 & U_{11} G_i
U_{14}^{\dagger} & 0  \\ \hline \hline U_{12} G_i U_{11}^{\dagger} &
U_{12} G_i U_{12}^{\dagger} & 0 & U_{12} G_i U_{13}^{\dagger} & 0 &
U_{12} G_i U_{14}^{\dagger} & 0 \\ \hline 0 & 0 & (G_i)_N & 0 & 0 &
0 & 0 \\ \hline \hline U_{13} G_i U_{11}^{\dagger} & U_{13} G_i
U_{12}^{\dagger} & 0 & U_{13} G_i U_{13}^{\dagger} & 0 & U_{13} G_i
U_{14}^{\dagger} & 0 \\ \hline 0 & 0 & 0 & 0 & 0 & 0 & 0 \\
\hline \hline U_{14} G_i U_{11}^{\dagger} & U_{14} G_i
U_{12}^{\dagger} & 0 & U_{14} G_i U_{13}^{\dagger} & 0 & U_{14} G_i
U_{14}^{\dagger} & 0 \\ \hline 0 & 0 & 0 & 0 & 0 & 0 & 0
\end{array} \right) \label{a4}\eeq

If we ignore all the columns and rows which are trivially null and
paste only those strips in which it is possible to find entries $
\neq 0 $, as discussed before, we end up with a new square matrix of
dimension

\begin{eqnarray} d & = & \frac{ ( N+2 ) ( N+3 ) ( N+4 ) }{6} + \frac{ ( N+2 ) (
N+3 ) }{2} + ( N+2 ) + 1 \nonumber \\ & = & \frac{ ( N+3 ) ( N+4 ) (
N+5 ) }{6} \label{a5} \end{eqnarray}

which is the dimension of the irreducible representation $ ( G_i
)_{N+2} $. Verifying that we obtain exactly the same values of the
representation $ ( G_i )_{N+2} $ is rather tedious. Let's briefly
check the case of $ G_5 $, whose representation is a diagonal
matrix,

\beq G_5 = \sum_{n_1,n_2,n_3,n_4 = 0}^{\infty} ( n_1 + n_2 - n_3 -
n_4 ) | n_1, n_2, n_3, n_4 >< n_1, n_2, n_3, n_4 | \label{a6} \eeq

and therefore we must control only the diagonal terms

\begin{eqnarray}
U_{11} G_5 U_{11}^{\dagger} & = & \sum_{n_1,n_2,n_3,n_4 =
0}^{\infty} ( n_1 + n_2 + 1 - n_3 - n_4 ) | n_1, n_2, n_3, n_4 ><
n_1, n_2, n_3,
n_4 | \nonumber \\
U_{12} G_5 U_{12}^{\dagger} & = & \sum_{n_1,n_2,n_3,n_4 =
0}^{\infty} ( n_2 + 1 - n_3 - n_4 ) | n_2, n_3, n_4, 0  ><  n_2,
n_3,
n_4, 0 | \nonumber \\
U_{13} G_5 U_{13}^{\dagger} & = & \sum_{n_1,n_2,n_3,n_4 =
0}^{\infty} ( - n_3 - n_4 - 1 ) | n_3, n_4, 0, 0 >< n_3,
n_4, 0 , 0 | \nonumber \\
U_{14} G_5 U_{14}^{\dagger} & = & \sum_{n_1,n_2,n_3,n_4 =
0}^{\infty} ( - n_4 - 1 ) | n_4, 0, 0, 0, 0 >< n_4, 0, 0, 0 |
\label{a7} \end{eqnarray}

Let's consider firstly the last term which is the simplest to
discuss. In fact restricting it to a fixed total oscillator number $
( N+1 ) $, there is only one element left

\beq ( U_{14} G_5 U^{\dagger}_{14} )_{N+1} = - ( N+2 ) | N+1, 0, 0,
0 >< N+1, 0, 0, 0 | \label{a8} \eeq

whose value $ - ( N+2 ) $ corresponds to the minimum value of the
representation $ ( G_5 )_{N+2} $. Studying in detail such
representation we notice that this minimum term is repeated $ (N+3)$
times, so we must find other $ (N+2)$ copies of it. It is not
difficult to realize that all those terms come from the restriction
of

\beq ( U_{13} G_5 U^{\dagger}_{13} )_{N+1} = - ( N+2 )
\sum_{n_3,n_4}^{n_3+n_4 = N+1} | n_3, n_4, 0, 0
>< n_3, n_4, 0, 0 | \label{a9} \eeq

Proceeding this way, we can verify that the terms $ ( U_{1i} G_5
U^{\dagger}_{1i} )_{N+1} \ ( i = 1,2 ) \ $ never reach the minimum
value $ - ( N+2 ) $ and it instead contribute to complete the
representation $ ( G_5 )_{N+2} $. With some patience the cases $ (
G_{1 \pm } )_{N+2} $ and $ ( G_{2 \pm } )_{N+2} $ can be
successfully checked.


\begin{thebibliography}{999}

\bibitem{1} Y. Kimura, " Noncommutative gauge theory on fuzzy
four-sphere and matrix model ", Nucl. Phys. {\bf B637} (2002) 177,
hep-th/0204256.
\bibitem{2} G. Landi, " Deconstructing monopoles and instantons ",
math-ph/9812004.
\bibitem{3} G. Landi, " Projective modules of finite type and
monopoles over $S^2$ ", J. Geom. Phys. {\bf 37} (2001) 47,
math-ph/9905014.
\bibitem{4} P. Valtancoli, " Projectors for the fuzzy sphere ",
Mod. Phys. Lett. {\bf A16} (2001) 639, hep-th/0101189.
\bibitem{5} P. Valtancoli, " Projective modules over the fuzzy four
sphere.", Mod. Phys. Lett. {\bf A17} (2002) 2189, hep-th/0210166.
\bibitem{6} S. Baez, A. P. Balachandran, B. Ydri, S. Vaidya, "
Monopoles and solitons in fuzzy physics ", Comm. Math. Phys. {\bf
208} (2000) 787, hep-th/9811169.
\bibitem{7} M. Dubois-Violette, Y. Georgelin, " Gauge theory in
terms of projector valued fields ", Phys. Lett. {\bf B82} (1999)
251.
\bibitem{8} T. Eguchi, P. B. Gilkey, A. J. Hanson, " Gravitation,
gauge theories, and differential geometry ", Phys. Rept. {\bf 66}
(1980) 213.
\bibitem{9} H. Grosse, C. Klimcik and P. Presnajder, " On finite
4D quantum field theory in noncommutative geometry ", Comm. Math.
Phys. {\bf 180} (1996) 429, hep-th/9602115.
\bibitem{10} J. Castelin, S. Lee and W. Taylor, " Longitudinal
5-branes as 4-spheres in matrix theory ", Nucl. Phys. {\bf B256}
(1998) 334, hep-th/9712105.
\bibitem{11} N. R. Constable, R. C. Myers and O. Tafjord, "
Non-abelian Brane intersections ", JHEP {\bf 0106} (2001) 023,
hep-th/0102080.
\bibitem{12} P. Valtancoli, " Projectors, matrix models and
noncommutative monopoles ",Int. J. Mod. Phys. {\bf A19} (2004), 4641
, hep-th/0404045.
\bibitem{13} P. Valtancoli, " Projectors, matrix models and
noncommutative instantons ", Int. J. Mod. Phys. {\bf A19} (2004),
4789, hep-th/0404046.
\bibitem{14} P. M. Ho, S. Ramgoolam, " Higher dimensional
geometries from matrix brane constructions ", Nucl. Phys. {\bf B627}
(2002), 266; hep-th/0111278.
\bibitem{15} P. Valtancoli, " $U(2)$ projectors and 't Hooft -
Polyakov monopoles on a fuzzy sphere ", hep-th/0511034.

\end{thebibliography}
\end{document}